\documentclass{article}

\usepackage[english]{babel}
\usepackage{latexsym}
\usepackage{amsmath}
\usepackage{dsfont}
\usepackage{bbm}
\usepackage{amssymb}
\usepackage{amsfonts}
\usepackage{mathbbol}
\usepackage{oldgerm}
\usepackage{euscript}
\usepackage[dvips]{graphicx}

\newtheorem{defi}{Definition}[section]

\newtheorem{theo}[defi]{Theorem}

\newtheorem{assump}[defi]{Assumption}

\def\R{\mathbb{R}}
\def\Mi{\mathbb{M}}
\def\N{\mathbb{N}}

\def\C{\mathbb{C}}

\def\1{\mathbbm{1}}

\def\Aa{\mathfrak{A}}
\def\Ba{\mathfrak{B}}

\def\Ma{\mathfrak{M}}
\def\Na{\mathfrak{N}}

\def\Cg{\mathcal{C}^\infty}

\def\D{\mathcal{D}}

\def\F{\mathcal{F}}

\def\H{\mathcal{H}}
\def\L{\mathcal{L}}
\def\M{\mathcal{M}}
\def\O{\mathcal{O}}
\def\P{\mathcal{P}}

\def\V{\mathcal{V}}
\def\W{\mathcal{W}}

\def\o{\omega}

\def\*{{}^{*}}

\def\supp{{\text{supp}}}

\def\proof{{\hspace{-0.7cm} Proof: }}

\def\qed{{\begin{flushright}$\Box$\end{flushright}}}

\begin{document}



\title{On the Generator of Massive Modular Groups}

\author{Timor Saffary\\
Wirtschafts- und Organisationswissenschaften\\ Helmut-Schmidt-Universit\"at\\Universit\"at der Bundeswehr Hamburg\\ D-22043 Hamburg\\ \small{e-mail: saffary@hsu-hh.de}}

\maketitle

\begin{abstract}
The purpose of this paper is to shed more light on the transition from the known massless modular action to the wanted massive one in the case of forward light cones and double cones. The infinitesimal generator $\delta_m$ of the modular automorphism group $\big(\sigma_m^t\big)_{t\in\R}$ is investigated, in particular, some assumptions on its structure are verified explicitly for two concrete examples.
\end{abstract}










\section{Introduction}

Modular Theory or Tomita-Takesaki theory is the generalisation of the modular function, which constitutes the difference between the left and right Haar measure, to non-commutative algebras. Although the prerequisite for this theory is only the specification of an underlying von Neumann algebra $\Ma$ and a cyclic and separating vector $\Omega\in\H$ or, equivalently, a faithful and normal state $\o$, it provides a deep insight into the most complex structure of von Neumann algebras. The main properties of the modular objects are addressed in Tomita's theorem \cite{Takesaki:1970}, i.e., the anti-unitary modular conjugation $J$ relates $\Ma$ to its commutant $\Ma'$, 
\begin{gather*}
J\Ma J=\Ma',
\end{gather*}
and the positive, selfadjoint modular operator $\Delta$ ensures the existence of an automorphism group on $\Ma$, 
\begin{align*}
\sigma_\o^t(A):=\pi_\o^{-1}\big(\Delta^{it}\pi_\o(A)\Delta^{-it}\big),\quad A\in\Ma,
\end{align*}
for all $t\in\R$, where $\pi_\o$ is the cyclic GNS representation of $\Ma$ with respect to the faithful state $\o$. These statements, in particular that a state already determines the dynamics of a system, have far-reaching consequences in mathematics as well as in physics. 

To start with, Connes shows that modular groups are equivalent up to inner automorphisms, i.e., two arbitrary groups $\sigma_{\o_1}^t$ and $\sigma_{\o_2}^t$ with respect to the states $\o_1$ and $\o_2$, respectively, are linked via a one-parameter family of unitaries $\Gamma_t$, the so-called cocycle,
\begin{gather}\label{cocycle}
\sigma_{\o_2}^t(A)=\Gamma_t\sigma_{\o_1}^t(A)\Gamma_t^*,\quad\forall A\in\Ma,t\in\R.
\end{gather}
This suggests the introduction of the modular spectrum $S(\Ma):=\bigcap_\o\mathbf{Spec}\Delta_\o$ by means of which Connes gives a complete classification of factors \cite{Connes:1973}, i.e., von Neumann algebras with $\Ma\cap\Ma'=\C\1$: 
\begin{itemize}
\item[$\bullet$] $\mathfrak{M}$ is of type $I$ or type $II$, if $S(\mathfrak{M})=\{1\}$;
\item[$\bullet$] $\mathfrak{M}$ is of type $III_0$, if $S(\mathfrak{M})=\{0,1\}$;
\item[$\bullet$] $\mathfrak{M}$ is of type $III_\lambda$, if $S(\mathfrak{M})=\{0\}\cup\{\lambda^{n}|\;0<\lambda<1,n\in \mathbb{Z}\}$;
\item[$\bullet$] $\mathfrak{M}$ is of type $III_1$, if $S(\mathfrak{M})=\R_+$.
\end{itemize}
The next development of paramount transboundary importance is Jones' classification of type $II_1$ subfactors \cite{Jones:1983kv}. He shows, contrary to everyone's expectation, that for the (global) index $[\Ma:\Na]$ not all positive real numbers are realised, but
\begin{gather*}
[\Ma:\Na]\in\Big\{4\cos^2\frac{\pi}{n}|\;n\in\N,n\geq3\Big\}\cup\big[4,\infty\big].
\end{gather*}
This result is extended by Kosaki to arbitrary factors \cite{Kosaki:1986}. Jones' index theory on his part connects widely separated areas, such as parts of statistical mechanics with exactly solvable models, and leads to some groundbreaking developments, e.g., a new polynomial invariant for knots and links in $\R^3$.

The interplay of modular theory and quantum field theory is most naturally apparent in the algebraic formulation since here the requirements of modular theory are already fulfilled: an underlying von Neumann algebra $\Ma(\O)$ is given and, due to the Reeh-Schlieder theorem, a cyclic and separating vacuum vector.

The usual formulation of quantum field theory is based on the representation of states as unit rays in a Hilbert space, with observables as operators acting on them. The algebraic ansatz proceeds in the opposite direction. There, the starting point are observables as elements of an abstract $\*$-algebra, on which the states are introduced as normalised, positive and linear functionals. For more details we refer the reader to the main source \cite{Haag:1963dh}, but also to \cite{Haag:1992hx} and \cite{Araki:1999ar}.

First one constructs a net of $C^*$-algebras $\{\Aa(\O)\}_{\O\subset\M}$, the so-called local algebras, i.e., to each open subset $\O$ of the spacetime $\M$ a $C^*$-algebra $\Aa(\O)$ is assigned that represents physical quantities to be measured in $\O$,
\begin{equation}\label{net}
\O\mapsto\Aa(\O).
\end{equation}
Since all physical information is assumed to be encoded in this mapping, its knowledge allows one in principle to extract all kinds of physical data. The $C^{*}$-Algebra $\Aa:=\overline{\bigcup_{\O}\Aa(\O)}$, i.e., the $C^*$-inductive limit of the net $\{\Aa(\O)\}_{\O\subset\M}$, is called the quasi-local algebra of observables and the bicommutant $\mathfrak{A}''$ of $\Aa$ the global algebra of observables. Two nets of local observables, $\Aa(\O)$ and $\tilde{\Aa}(\O)$, respectively, are said to be mutually isomorphic if there is an isomorphism $i:\Aa\rightarrow\tilde{\Aa}$  with  $i[\Aa(\O)]=\tilde{\Aa}(\O)$. In addition the net is required to satisfy the following conditions:

\begin{itemize}
\item[(i)] Isotony: $O_1\subset\mathcal{O}_{2}\;\Longrightarrow\;\Aa(\O_1)\subset\Aa(\O_2)$.    
\item[(ii)] Locality: $[\Aa(\O_1),\Aa(\O_2)]=\{0\}$ for spacelike separated $\O_1$ and $\O_2$.
\item[(iii)] Additivity: $\O=\underset{i}{\bigcup}\O_i\;\Longrightarrow\;\Aa(\O)=\Big(\underset{i}{\bigcup}\Aa(\O_i)\Big)''$.
\item[(iv)] Covariance: There is a strongly continuous unitary representation $U(\P)$ of the Poinc\'are group $\P$, $\Aa(g\O)=U(g)\Aa(\O)U(g)^{-1},g\in\P$.
\item[(v)] Spectrum condition: $\mathbf{Spec}\,U(g)\subseteq\V_+$.
\item[(vi)] Vacuum sector: There exists a vector $\Omega\in\H,\;\|\Omega\|=1$, such that $U(g)\Omega=\Omega,\;g\in\P$, and $\Big(\underset{\O\subset\M}{\bigcup}\Aa(\O)\Big)\Omega$ is dense in $\H$.
\end{itemize}
If  $\M=\mathbb{M}$ is the Minkowski space, then the isometries turn out to be the Poincar\'e transformations, and the last condition becomes the Poincar\'e covariance and cyclicity of the vacuum vector $\Omega$. A state on the observable algebra $\Aa$ is represented by a linear functional $\o :\Aa\longrightarrow\C$, which is normalised, i.e., $\omega(\1)=1$, and positive, i.e. $\o(A^*A)\geq 0$ for all $A\in\Aa$. The `usual' and the algebraic formulation of quantum field theory can be connected via the GNS representation.
 
The first physical application of modular theory is proved by Takesaki who recognises that the equilibrium dynamics is determined by the modular groups, since their infinitesimal generator is the thermal Hamiltonian and they satisfy the KMS condition, the generalisation of Gibbs' notion of equilibrium to systems with infinitely many degrees of freedom, 
\begin{gather*}
\o\big(A\sigma_\o^{i\beta}(B)\big)=\o(BA), 
\end{gather*}
where $\beta$ is the inverse of the temperature. 

The classification theory is not less important in physics, in fact, the quest for decomposition of quantum systems has been one of von Neumann's most important reasons for the investigation of operator algebras. In local quantum physics, one is interested in the structure of the von Neumann algebra of local observables $\Ma(\O)$. The analysis, which has been undertaken by a colloboration of many persons, discovered $\Ma(\O)$ as a hyperfinite factor of type $III_1$. The substructure of $\Ma(\O)$, which is of utmost significance for decoding the physical information contained in the mapping \eqref{net}, is determined only for conformal local nets with central charge $c<1$ yet \cite{Kawahigashi:2002px}.

\begin{figure}[here]\label{grafik1}
\begin{center}
 \includegraphics[height=3cm]{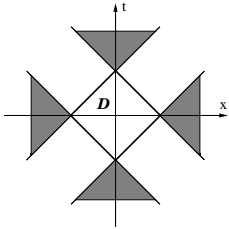}
    \caption[]{ \footnotesize \em Geometric action of $J_\D$}
\end{center}
\end{figure}

The third main application for modular theory in local quantum physics is the modular action as a geometric transformation on the local algebra for special spacetimes. For the local algebra generated by Wightman fields with mass $m\geq0$ localised in the right wedge $\W_R:=\big\{x\in\mathbb{M}|\;|x^0|<x^1\big\}$ Bisognano and Wichmann identify  the modular action with the Lorentz boosts $\Lambda$ and the modular conjugation with the TCP operator $\Theta$ \cite{Bisognano:1975ih},\cite{Bisognano:1976za}:
\begin{gather*}
J_{\W_R}=\Theta U\big(R_1(\pi)\big),\quad J_{\W_R}\Ma(\W_R)J_{\W_R}=\Ma(\W_L),\\
\sigma_{\W_R}^t\big(\varphi[f]\big)=\varphi[f_s],\quad f_s(x):=f\big(\Lambda_{s}(x)\big),\quad x\in\W_R,\; s:=2\pi t,
\end{gather*}
where $R_1$ denotes the spatial rotation around the $x^1$-axis and $\W_L:=\big\{x\in\mathbb{M}|\;|x^0|<-x^1\big\}$ is the left wedge. For massless theories this result has been transferred via conformal transformations to other spacetime regions. For forward light cones $\V_+:=\big\{x\in\mathbb{M}|\;x\cdot x>0\text{ and } x^0>0\big\}$ the modular action conicides with dilations and $J_{\V_+}$ maps $\V_+$ onto the backward light cone $\V_-:=\big\{x\in\mathbb{M}|\;x\cdot x>0\text{ and } x^0<0\big\}$ as shown by Buchholz \cite{Buchholz:1977ze}. For double cones $\D:=\V_+\cap\V_-$, $J_\D$ maps $\D$ onto the shaded region in Figure \ref{grafik1},

\hspace*{-7,5mm} while the modular group acts as conformal transformations as proved by Hislop and Longo \cite{Hislop-Longo:1981uh}, $x\in\D,s\in\R$:
\begin{gather*}
x_\pm(s)=\frac{1+x_\pm-e^{-s}(1-x_\pm)}{1+x_\pm+e^{-s}(1-x_\pm)},\quad x_\pm:=x^0\pm|\mathbf{x}|.
\end{gather*}
The geometric interpretation of the modular group is of paramount importance in further applications. The result of Bisognano and Wichmann is not only closely related to the Unruh effect and the black hole evaporation, actually, in analogy of the Rindler wedge with a black hole, it implies the Hawking radiation, but has also made possible the derivation of some most fundamental concepts of quantum field theory, as there are the proofs of the PCT theorem by Borchers \cite{Borchers:1991xk} and of the spin-statistics theorem by Guido and Longo \cite{Guido:1995fy}, where modular theory intervenes twice through Jones' index theory, the construction of the Poincar\'e group by Brunetti, Guido and Longo \cite{Brunetti:1992zf}, and the introduction of modular nuclearity condition by Buchholz, D'Antoni and Longo \cite{Buchholz:1989bj}, nuclearity as a tool to single out models with decent phase space properties. Moreover, Schroer and Wiesbrock's investigation gives a hint that modular theory plays a decisive r\^ole in the construction of field theories with interaction \cite{Schroer:1998wp}. 

As aforementioned, the modular group for massive theories, $\sigma_m^t$, is known only for wedge regions. In fact, the transfer of Bisognano and Wichmann's result via conformal mappings to forward light cones and double cones does not work in massive theories. If one assumes the modular group to act locally, then, as shown by Trebels \cite{Trebels:1997}, the action can be determined as the ones of Bisognano-Wichmann, Buchholz and Hislop-Longo up to a scaling factor. But in general the modular action has to be non local and does not act as a  geometric transformation anymore. This is mainly due to the fact that the massive scalar field is not invariant under conformal transformations. 

Modular theory in general seems to become more and more a powerful tool for diverse problems and the natural formalism by which local quatum physics may be formulated. But the potential of modular theory will not be exhausted fully as long as the modular group $\sigma_m^t$ acting on the massive algebras $\Ma_m(\O)$ is not determined. With $\sigma_m^t$ one would obtain a deeper and easier accessible understanding of the dynamics of quantum systems.

More details on the state-of-the-art of the applications of modular theory in mathematics and physics as well as on the results given below can be obtained in \cite{Saffary:2005mz}.

\section{Why a Pseudo-Differential Operator?}

Since the discovery of Bisognano and Wichmann, there have been many attempts to derive the massive modular group in the case of double cones, the most important spacetime regions, but no progress has been made so far. What should be mentioned are some assumptions on its nature, to be more precise, on the structure of its infinitesimal generator $\delta_m$, where $m$ denotes the mass. It is well known that the generators $\delta_0$ of the massless groups are all ordinary differential operators of order one. Because of the non local action and a result of Figliolini and Guido, who prove that $\delta_m$ is depending on $m$ in the strong generalised sense \cite{Figliolini:1989vf}, one assumes that the generator $\delta_m$ has to be a pseudo-differential operator (PsDO), the generalisation of differential operators. 

The transition to PsDOs is easily seen, if we express an arbitrary differential operator with variable coordinates,
\begin{gather*}
p(x,D):=\sum_{|\alpha|\leq m}a_{\alpha}(x)D_{x}^{\alpha},\quad a_{\alpha}\in \Cg(\Omega),\quad D_x^{\alpha}:=(-i)^{|\alpha|}\partial_x^{\alpha},
\end{gather*}
with $u,f\in\mathcal{D}'(\Omega)$, $\Omega\subset\R^{n}$, with the following form,
\begin{equation}\label{PsDO-defi}
\begin{split}
p(x,D)u(x)&=\frac{1}{(2\pi)^{\frac{n}{2}}}p(x,D)\int_{\R^{n}}\tilde{u}(\xi)e^{ix\xi}d^{n}\xi\\
&=\frac{1}{(2\pi)^{\frac{n}{2}}}\int_{\R^{n}}p(x,\xi)\tilde{u}(\xi)e^{ix\xi}d^{n}\xi,
\end{split}
\end{equation}
where $\tilde{u}$ denotes the Fourier transform of $u$ and $p(x,\xi):=\sum_{|\alpha|\leq m}a_{\alpha}(x)\xi^{\alpha}$. This concept of differential operators with variable coordinates can be generalised by replacing the polynomial $p(x,\xi)$ with the so-called symbols. 

\begin{defi}\label{symbol-definition}
Let $m\in\R$ and $0<\delta\leq\rho\leq 1,\,\delta<1$. Then $p\in\mathcal{C}^{\infty}(\R_{x}^{n}\times\R_{\xi}^{n})$ is said to be a symbol of order $m$ and of type $(\rho,\delta)$ if $p$ satisfies the following condition:
\begin{equation}
\begin{split}
\forall\alpha,\beta\in\N_{0}^{n}:\,\exists C_{\alpha,\beta}\geq 0:\;\Big(\big|\partial_{\xi}^{\alpha}\partial_{x}^{\beta}p(x,\xi)\big|\leq C_{\alpha,\beta}&\big(1+|\xi|\big)^{m+\delta|\beta|-\rho|\alpha|}\\
&\forall(x,\xi)\in\R_x^{n}\times\R_\xi^{n}\Big).
\end{split}
\end{equation}
\end{defi}
It should be mentioned that some authors use a different notion of symbols, $p\in S_{\rho,\delta}^{m}(K\times\R_{\xi}^{n})$, often referred as H\"ormander's version, by demanding the variable $x$ to lie in a compact subset $K\subset\Omega$, where $\Omega$ is an open subset of $\R_x^n$. But be aware that H\"ormander works with both versions, with the latter definition in \cite{Hoermander:1971} and with Definition \ref{symbol-definition} in \cite{Hoermander:1990}.\\ 
One may generalise PsDOs by addmitting in \eqref{PsDO-defi}, instead of the scalar product $x\cdot\xi$, general phase functions $\theta(x,\xi)$ in Equation \eqref{PsDO-defi} which satisfy the following conditions:
\begin{itemize}
\item[(i)] $\theta(x,\xi)$ is (in general) complex-valued, smooth and homogeneous of degree one. 
\item[(ii)] The gradient $\nabla_x\theta(x,\xi)$ does not vanish on the conic support of the symbol $a(x,\xi)$ for $\xi\neq 0$.
\end{itemize}
This leads to the definition of so-called Fourier integral operators,
\begin{gather}\label{FIO}
Au(x):=\int a(x,\xi)\tilde{u}(\xi)e^{i\theta(x,\xi)}d\xi.
\end{gather}
Now, we are able to state our conjecture on the massive generator.

\begin{assump}\label{assumption-gen0}
 The massive infinitesimal generator $\delta_m$ has the following structure:
\begin{gather}\label{assumption-gen1}
\delta_m=\delta_0+\delta_r,
\end{gather}
where $\delta_0$ is the massless generator, satisfying the following properties:
\begin{itemize}
\item[(i)] The known massless generator $\delta_0$ is the principal term in $\delta_m$.
\item[(ii)] $\delta_r$ is a PsDO.
\end{itemize}
\end{assump}

This assumption  is mentioned explicitly first by Schroer and Wiesbrock \cite{Schroer:1998ax}. However, the authors restrict themselves to a few remarks on their strategy how to calculate $\delta_m$ without giving proofs. This strategy supposes an intermediate step, namely the computation of the modular group $\sigma_{m,0}^t$ on $\Ma_m(\D)$ with respect to the `wrong' massless vacuum vector $\Omega_0$.  They propose to derive, in a final step, the modular group $\sigma_m^t$ and its generator $\delta_m$ with respect to the `right' massive vacuum vector $\Omega_m$ via cocycle theorem \eqref{cocycle}. They assume that this procedure will not change the pseudo-differential nature of the infinitesimal generator.

\section{Modular Groups with Non Local Action}

We want to investigate the Assumption \ref{assumption-gen0} on its infinitesimal generator $\delta_m$, which is obviously not known in general. Due to the result of Trebels, $\sigma_m^t$ has to act non-locally on $\Ma_m(\O)$, otherwise it would coincide with $\sigma_0^t$ up to a scaling factor. Thus, examples of modular automorphism groups acting non-locally may serve as a testing ground for the aforementioned assumptions on their infinitesimal generators $\delta_m$. To the best of our knowledge, there exist only two concrete examples for such modular groups in the literature. The first modular automorphism group is given by Yngvason in the context of his investigation on essential duality \cite{Yngvason:1994nk}, and the second one is introduced by Borchers and Yngvason who formulate modular groups in a more general setting, namely with respect to arbitrary KMS states instead of the vacuum state \cite{Borchers:1998ye}.

\subsection{Yngvason's Counter-example}

In a Poincar\'e covariant Wightman framework, Bisognano and Wichmann identify the modular groups with the Lorentz boosts and, furthermore, show that wedge duality holds. Yngvason investigates the validity of these two properties for local nets \cite{Yngvason:1994nk}. He can give explicit examples for fields violating essential duality, an implication of wedge duality and major assumption in the superselection theory, or Lorentz covariance. We take one of his concrete examples as an opportunity to analyse the infinitesimal generator of a modular group with nonlocal action.

 He starts with a Hermitian Wightman field $\varphi$ which transforms covariantly under spacetime translations, but not necessarily under Lorentz transformations. For the sake of simplicity let us consider their special two-point function consisting of only one term,
$$
\o_2(p)=M(p)d\mu(p),
$$
whose polynomial factorises as
$$
M(p)=:F(p)F(-p)\quad\text{and}\quad F(p)^*=F(-p),
$$
where $F(p)$ (in general no polynomial) is analytic and has no zeros in the right wedge characterised by $x_+>0$ and $ x_-<0$. The existence of such polynomials is ensured by the following example,
\begin{gather}
M(p):=\sum_{i=1}^n(p^i)^2+m^2,\quad F(p)=(\hat{p}\hat{p}+m^2)^{1/2}+\frac{i}{2}(p_++p_-),\label{Yngvason-example}
\end{gather}
where we have used the notation $\hat{p}:=(p^2,\cdots,p^n)$ and $p_\pm:=p^0\pm p^1$. One obtains the generalised free field $\partial_t\varphi_m(x)$, where $\varphi_m$ is the free field of mass $m$, by setting $d\mu(p):=\Theta(p^0)\delta\big((p,p)_\Mi-m^2\big)$ and $M(p):=(p^0)^2$. For $\lambda>0$ one can now define the unitary operator $V_{\W_R}(\lambda)$ on the Fock space $\F$, first on the one-particle space $\H_1:=\L^2\big(\R^n,M(p)d\mu(p)\big)$ by
\begin{equation*}
V_{\W_R}(\lambda)\varphi(p):=\frac{F(-\lambda p_+,-\lambda^{-1} p_-,-\hat{p})}{F(-p_+,-p_-,-\hat{p})}\varphi(\lambda p_+,\lambda^{-1} p_-,\hat{p})
\end{equation*}
for $\varphi\in\H_1$, and then by canonical extension (second quantisation) to $\F$. One then introduces a one-parameter group of automorphisms on the von Neumann algebra $\mathfrak{M}(\W_R)$ over $\H$ generated by the Weyl operators $W(f):=e^{i\varphi[f]}$:
\begin{equation}\label{Yngvason}
\sigma_{\W_R}^t\big(W(f)\big):=V_{\W_R}\big(e^{-2\pi t}\big)W(f)V_{\W_R}\big(e^{2\pi t}\big).
\end{equation}
Yngvason identifies this group with the modular group with respect to the vacuum state on $\mathfrak{M}(\W_R)$ by proving the validity of the KMS condition. Since the operator $V_{\W_R}(\lambda)$ maps the Fourier transform $\tilde{f}$ of $f$ with $\supp f\subset \W_R$ into
\begin{equation*}
V_{\W_R}(\lambda)\tilde{f}(p)=\frac{(\hat{p}\hat{p}+m^2)^{1/2}-\frac{i}{2}(\lambda p_+-\lambda^{-1}p_-)}{(\hat{p}\hat{p}+m^2)^{1/2}-\frac{i}{2}(p_+-p_-)}\tilde{f}(\lambda p_+,\lambda^{-1}p_-,\hat{p}),
\end{equation*}
which is not analytic in $\hat{p}$ and therefore cannot be the Fourier transform of a function with compact support in the $\hat{x}$-direction, $\hat{x}:=(x^2,\cdots,x^n)$, $W(f_\lambda)$ cannot be an element of any wedge algebra unless the wedge is a translate of $\W_R$ or the left wedge $\W_L:=\big\{x\in\mathbb{M}|\;|x^0|<-x^3\big\}$. The operator $W(f_\lambda)$ is still localised only in the $x^0,x^1$-directions in the sense that it is an element of $\mathfrak{M}(\W_R+a)\cap\mathfrak{M}(\W_R+b)'$ for some $a,b\in\W_R$.

One may ask if the non-local behaviour of this example is reflected in some way by the infinitesimal generator of the group \eqref{Yngvason}. First, we derive the generator for the modular group,
\begin{align*}
\Delta_{\W_R}^{it}\varphi(p)&=\frac{F(-\lambda p_+,-\lambda^{-1}p_-,-\hat{p})}{F(-p_+,-p_-,-\hat{p})}\varphi(\lambda p_+,\lambda^{-1}p_-,\hat{p}),
\end{align*}
where $\lambda=e^{-2\pi t}$, as
\begin{align*}
\delta_{\W_R}\varphi(p)&=\partial_t\Delta_{\W_R}^{it}\varphi(p)\big|_{t=0}\\
&=\bigg\{\frac{2\pi}{F(-p_+,-p_-,-\hat{p})}\big(p_+\partial_{p_+}-p_-\partial_{p_-}\big)F(-p_+,-p_-,-\hat{p})\\
&\qquad-2\pi p_+\partial_{p_+}+2\pi p_-\partial_{p_-}\bigg\}\varphi(p_+,p_-,\hat{p}).
\end{align*}
For our example \eqref{Yngvason-example} we obtain:
\begin{align}\label{Yngvason-generator}
\delta_{\W_R}\varphi(p)
&=\bigg\{\frac{-2i\pi p^1}{(\hat{p}\hat{p}+m^2)^{1/2}-ip^0}-4\pi\big(p^0\partial_{p^1}+p^1\partial_{p^0}\big)\bigg\}\varphi(p_+,p_-,\hat{p}).
\end{align}
While the second term in \eqref{Yngvason-generator} can be identified with the Bisognano-Wichmann infinitesimal generator, the first term containing the mass $m$ is a PsDO of order zero. This additional part has to comprise the non local character of the modular group $\Delta_{\W_R}^{it}$. To put it in a nutshell, we have verified the Assumption \ref{assumption-gen0} with 
\begin{gather}
\delta_r:=\frac{-2i\pi p^1}{(\hat{p}\hat{p}+m^2)^{1/2}-ip^0}.
\end{gather}

\subsection{Borchers-Yngvason's Counter-example}

In \cite{Borchers:1998ye} Borchers and Yngvason give other examples for modular automorphism groups which act non locally on the wedges, light cones and double cones. Whereas all investigations given so far have been concerned with modular groups with respect to the vacuum state, Borchers and Yngvason formulate the automorphism groups by means of KMS states.

They start with a general $C^*$-dynamical system $(\Aa,\alpha^t)$, an $\alpha^t$-invariant subalgebra $\Ba$, i.e., $\alpha^t(\Ba)\subseteq\Ba$, and an $(\alpha,\beta$)-KMS state $\o$. Due to the analyticity property of KMS states, $\Omega$ is separating, and also cyclic for $\Ma:=\pi_\o(\Aa)''$ and $\Na:=\pi_\o(\Ba)''$, if one assumes $\bigcup_{t\in\R}\alpha^t(\Ba)$ to be dense in $\Aa$ in the norm topology. Hence, the existence of the modular objects is ensured and one may determine the action of the modular automorphism group. The authors restrict themselves to two-dimensional theories which factorise in the light cone variables $x_\pm:=x^0\pm x^1$. In these cases one may first establish the modular group on the algebra $\Ma(\R_+)$ as
\begin{gather*}
\Delta_+^{i\tau}\Ma\big([x_\pm,\infty[\big)\Delta_+^{-i\tau}=\Ma\big([\nu_+^t(x_\pm),\infty[\big),\\
\nu_+^t(x_\pm):=\frac{\beta}{2\pi}\log\Big(1+e^{-2\pi t}\big(e^{2\pi x_\pm/\beta}-1\big)\Big)
\end{gather*}
for all $t,x_\pm\in\R$ satisfying
\begin{gather}\label{Borchers-Yngvason2}
1+e^{-2\pi t}\big(e^{2\pi x_\pm/\beta}-1\big)>0.
\end{gather}
In the same manner one introduces the modular group on the algebra $\Ma(\R_-)$ as
\begin{gather*}
\Delta_+^{i\tau}\Ma\big(]-\infty,x_\pm]\big)\Delta_+^{-i\tau}=\Ma\big(]-\infty,\nu_-^t(x_\pm)]\big)
\end{gather*}
with $\nu_-^t(x_\pm):=-\nu_+^{-t}(-x_\pm)$ for all $t,x_\pm\in\R$ fulfilling
\begin{gather}\label{Borchers-Yngvason3}
1+e^{2\pi\tau}\big(e^{-2\pi x_\pm/\beta}-1\big)>0.
\end{gather}
Now, one can express the algebra for the two-dimensional space $I_+\times I_-\subseteq\R^2$ via the tensor product $\Ma(I_+\times I_-)=\Ma(I_+)\otimes\Ma(I_-)$, in particular, one obtains for the examples of our interest:
\begin{gather*}
\Ma(\W_R)=\Ma(\R_-)\otimes\Ma(\R_+),\quad\Ma(\V_+)=\Ma(\R_+)\otimes\Ma(\R_+),\quad\\
\text{and}\quad\Ma(\O)=\Ma(I_-)\otimes\Ma(I_+).
\end{gather*}
The corresponding modular groups with respect to a factorising KMS state $\o\otimes\o$ are given  in the case of $\V_+$ as:
\begin{gather*}
\Delta_{\V_+}^{it}\varphi[f]\Delta_{\V_+}^{-it}=\varphi\big[\nu_{\V_+}^tf\big],\quad\big(\nu_{\V_+}^tf\big)(x_-,x_+):=f\big(\nu_+^t(x_-),\nu_+^t(x_+)\big),
\end{gather*}
for all $t\in\R$ and $x_\pm\in\R$ satisfying \eqref{Borchers-Yngvason2} for $x=x_\pm$. Close to the apex of $\V_+$, one obtains the known case $\beta=\infty$, i.e., dilations with the light cone coordinates $x_+$ and $x_-$ scaled by the factor $e^{-2\pi t}$. Analoguously, for the right wedge one has
\begin{gather*}
\Delta_{\W_R}^{i\tau}\varphi[f]\Delta_{\W_R}^{-i\tau}=\varphi\big[\nu_{\W_R}^tf\big],\quad\big(\nu_{\W_R}^tf\big)(x_-,x_+):=f\big(\nu_-^t(x_-),\nu_+^t(x_+)\big),
\end{gather*}
for all $t\in\R$ and $x_\pm\in\R$ satisfying \eqref{Borchers-Yngvason3} and \eqref{Borchers-Yngvason2} for $x=x_-$ and $x=x_+$, respectively. Here, near the edge of the the wedge, the action may be identified with the case $\beta=\infty$, i.e., with Lorentz boosts where the light cone coordinates $x_+$ and $x_-$ are scaled by the factors $e^{-2\pi t}$ and $e^{2\pi t}$, respectively.

For more concrete calculations Borchers and Yngvason investigate the Weyl algebra of free Bose fields generated by elements $W(f)$, $f\in\D(\R)$, with
\begin{gather*}
W[f]^*=W[-f],\quad W[f]W[g]=e^{-K(f,g)/2}W[f+g],\\
\text{and}\quad K(f,g):=\int_{-\infty}^\infty p\,Q(p^2)\tilde{f}(-p)\tilde{g}(p)dp,
\end{gather*}
where $Q(p^2)$ is a non-negative polynomial. They introduce for each scaling dimension $n\in\N$ and interval $I\subset\R$ the algebra $\Ma^{(n)}(I)$ which is generated by the Weyl operators $W^{(n)}[f]$ corresponding to $Q(p^2)=p^{2n}$. While the algebra is known to be independent of $n$ for unbounded  $I$, for bounded intervals one only has the inclusion 
\begin{gather}\label{Borchers-Yngvason-inclusion}
\Ma^{(m)}(I)\subset\Ma^{(n)}(I),
\end{gather}
whenever $m>n$. Thus the modular operators $\Delta_+$ and $\Delta_-$ corresponding to the positive real axis and the negative one, respectively, are independent of $n$. 

\begin{theo}[Borchers-Yngvason]\label{Borchers-Yngvason-theo3}
Let $\o$ be a quasi-free KMS state on the Weyl algebra $\Ma^{(0)}(\R_+)$ and $\pi$ the corresponding cyclic representation, then one has:
\begin{equation}
\begin{split}
\Delta_+^{it}\pi\big(W^{(0)}[f]\big)&\Delta_+^{-it}=\pi\big(W^{(0)}[\eta_+^{t,(0)}f]\big),\\
\big(\eta_+^{t,(0)}f\big)(x_\pm):=f\big(\nu_+^t(x_\pm)\big)=f&\left(\frac{\beta}{2\pi}\log\left\{1+e^{-2\pi t}\big(e^{2\pi x_\pm/\beta}-1\big)\right\}\right),
\end{split}
\end{equation}
with $\supp f\subset\R_+$. For $n>0$ one introduces:
\begin{equation}
\begin{split}
\Delta_+^{it}\pi\big(W^{(n)}[f]\big)&\Delta_+^{-it}=\pi\big(W^{(n)}[\eta_+^{t,(n)}f]\big),\\
\big(\eta_+^{t,(n)}f\big)(x_\pm):=\int_0^{x_\pm}\int_0^{x^1}\cdots&\int_0^{x^{n-1}}\eta_+^{t,(0)}f^{(n)}(x^n)dx^n\cdots dx^1,
\end{split}
\end{equation}
where $f^{(n)}$ is the $n$-th derivative of the test function $f$ with $\supp f\subset\R_+$.
\end{theo} 
The modular action on the negative axis is formulated as aforementioned, i.e., $\eta_-^{t,(0)}$ is defined via the transformation $\nu_-^t(x_\pm)$.

Borchers and Yngvason show that the modular group acts locally only in the case $n=0$. While the action on the field operator, which can be regained from the Weyl operators 
\begin{gather*}
\pi\big(W^{(n)}[f]\big)=:e^{i\int\varphi^{(n)}(x)f(x)dx}
\end{gather*}
through functional derivation, is
\begin{gather*}
\Delta_+^{it}\varphi^{(0)}(x_\pm)\Delta_+^{-it}=\partial_{x_\pm}\nu_+^t(x_\pm)\varphi^{(0)}\big(\nu_+^t(x_\pm)\big),
\end{gather*}
for $n=1$ one gets an additional term, e.g., at the origin
\begin{gather*}
\Delta_+^{it}\varphi^{(1)}(0)\Delta_+^{-it}=e^{-2\pi t}\varphi^{(1)}(0)-\frac{2\pi}{\beta}e^{-4\pi t}\int_0^\infty\varphi^{(1)}(x)dx.
\end{gather*}
In the case of double cones, namely where we are dealing with bounded intervals $I_\pm\subset\R_\pm$, fields of higher scaling dimension $\varphi^{(n)}$, $n\geq1$, are in general localised only in the algebra $\Ma^{(0)}(I_\pm)$ after the modular action, due to the inclusion \eqref{Borchers-Yngvason-inclusion}, but no longer in the original subalgebra $\Ma^{(0)}(I_\pm)$.

Also in this case, we are interested in the infinitesimal generator $\delta^{(n)}$ of the modular automorphism group acting on wedges, forward light cones and double cones, since we expect to see this non local behaviour in the pseudo-differential structure of $\delta^{(n)}$. The generator corresponding to the positive real axis in the case of $n=0$ is
\begin{gather*}
\delta_+^{(0)}\varphi^{(0)}[f]:=\partial_t\Delta_+^{it}\varphi^{(0)}[f]\Delta_+^{-it}\big|_{t=0}=\varphi^{(0)}\big[\partial_t\eta_+^{t,(0)}f\big]\big|_{t=0},\\
\big(\partial_t\eta_+^{t,(0)}f\big)(x_\pm)\big|_{t=0}=\partial_tf\big(\nu_+^t(x_\pm)\big)\big|_{t=0}=-\beta\big(1-e^{-2\pi x_\pm/\beta}\big)\partial_{x_\pm}f(x_\pm),
\end{gather*}
while the counterpart with respect to the negative real axis reads
\begin{gather*}
\delta_-^{(0)}\varphi^{(0)}[f]:=\partial_t\Delta_-^{it}\varphi^{(0)}[f]\Delta_-^{-it}\big|_{t=0}=\varphi^{(0)}\big[\partial_t\eta_-^{t,(0)}f\big]\big|_{t=0},\\
\big(\partial_t\eta_-^{t,(0)}f\big)(x_\pm)\big|_{t=0}=\partial_tf\big(\nu_-^t(x_\pm)\big)\big|_{t=0}=-\beta\big(1-e^{2\pi x_\pm/\beta}\big)\partial_{x_\pm}f(x_\pm).
\end{gather*}
In terms of the original spacetime coordinates the infinitesimal generators have the following form:
\begin{equation}\label{delta0}
\begin{split}
\delta_{\V_+}^{(0)}f(x^0,x^1)=\frac{\beta}{2}\Big[\big(&e^{-2\pi x_+/\beta}+e^{-2\pi x_-/\beta}-2\big)\partial_{x^0}\\
&+\big(e^{-2\pi x_+/\beta}-e^{-2\pi x_-/\beta}\big)\partial_{x^1}\Big]f(x^0,x^1),\\
\delta_{\W_R}^{(0)}f(x^0,x^1)=\frac{\beta}{2}\Big[\big(&e^{-2\pi x_+/\beta}+e^{2\pi x_-/\beta}-2\big)\partial_{x^0}\\
&+\big(e^{-2\pi x_+/\beta}-e^{2\pi x_-/\beta}\big)\partial_{x^1}\Big]f(x^0,x^1).
\end{split}
\end{equation}
The generator for an arbitrary $n>0$ is given in the next 
\begin{theo}\label{Borchers-Yngvason-gen}
The infinitesimal generator of the modular group acting on $\Ma^{(n)}(\R_+)$ is for $n>0$
\begin{equation}
\begin{split}
\delta_+^{(n)}f(x_\pm)&=\delta_+^{(n-1)}f(x_\pm)+\delta_{+,r}^{(n)}f(x_\pm),\\
\delta_{+,r}^{(n)}f(x_\pm):=2\pi&\int\frac{(i\xi)^n}{\big(i\xi-\frac{2\pi}{\beta}\big)^n}\tilde{f}(\xi)e^{ix_\pm(\xi+2\pi i/\beta)}d\xi.
\end{split}
\end{equation}
The counterpart for $\Ma^{(n)}(\R_-)$ for $n>0$ reads 
\begin{equation}
\begin{split}
\delta_-^{(n)}f(x_\pm)&=\delta_-^{(n-1)}f(x_\pm)+\delta_{-,r}^{(n)}f(x_\pm),\\
\delta_{-,r}^{(n)}f(x_\pm):=-2\pi&\int\frac{(i\xi)^n}{\big(i\xi+\frac{2\pi}{\beta}\big)^n}\tilde{f}(\xi)e^{ix_\pm(\xi-2\pi i/\beta)}d\xi.
\end{split}
\end{equation}
\end{theo}

\proof
By induction one obtains for the positive real axis:
\begin{align*}
\delta_+^{(n+1)}&f(x_\pm)=\partial_t\int_0^{x_\pm}\int_0^{x^1}\cdots\int_0^{x^n}\eta_t^{(0)}f^{(n+1)}(x^{n+1})dx^{n+1}\cdots dx^1\bigg|_{t=0}\\
&=\int_0^{x_\pm}\int_0^{x^1}\cdots\int_0^{x^n}\partial_tf^{(n+1)}\big(\nu_+^t(x^{n+1})\big)dx^{n+1}\cdots dx^1\bigg|_{t=0}\\
&=\int_0^{x_\pm}\int_0^{x^1}\cdots\int_0^{x^n}\partial_{x^{n+1}}^{n+2}f(x^{n+1})\partial_t\nu_+^t(x^{n+1})\big|_{t=0}dx^{n+1}\cdots dx^1\\
&=\int_0^{x_\pm}\int_0^{x^1}\cdots\int_0^{x^{n-1}}\partial_{x^n}^{n+1}f(x^n)\partial_t\nu_+^t(x^n)\big|_{t=0}dx^n\cdots dx^1\\
&\quad-\int_0^{x_\pm}\int_0^{x^1}\cdots\int_0^{x^n}\partial_{x^{n+1}}^{n+1}f(x^{n+1})\partial_{x^{n+1}}\partial_t\nu_+^t(x^{n+1})\big|_{t=0}\\
&\hspace{8cm}dx^{n+1}\cdots dx^1\\
&=\delta_+^{(n)}f(x_\pm)-\delta_{+,r}^{(n+1)}f(x_\pm).
\end{align*}
Due to the fact that $\supp f\subset\R_+$, we get for the additional term
\begin{align*}
\delta_{+,r}^{(n+1)}f(x_\pm)&=2\pi\int_0^{x_\pm}\int_0^{x^1}\cdots\int_0^{x^n}(i\xi)^{n+1}\tilde{f}(\xi)e^{ix_\pm^{n+1}\xi}e^{-2\pi x_\pm^{n+1}/\beta}\\
&\hspace{6cm}d\xi dx^{n+1}\cdots dx^1\\
&=2\pi\int\frac{(i\xi)^{n+1}}{\big(i\xi-\frac{2\pi}{\beta}\big)^{n+1}}\tilde{f}(\xi)e^{ix_\pm(\xi+2\pi i/\beta)}d\xi.
\end{align*}
The expression for the generator corresponding to $\R_-$ is calculated in the same way.\qed
\qed

What we have shown is that the infinitesimal generators $\delta_+^{(n)}$ and $\delta_-^{(n)}$ with scaling dimension $n\geq 1$ are no longer differential operators but Fourier integral operators instead, see Definition \eqref{FIO}. To be more precise, the generators do have the following structure:
\begin{gather*}
\delta_\pm^{(n)}=\delta_\pm^{(0)}+\sum_{k=1}^n\delta_\pm^{(k),r}=:\delta_\pm^{(0)}+\delta_{\pm,r}^{(n)}.
\end{gather*}
Whereas the principal symbol $\delta_\pm^{(0)}$ is still a differential operator of order one, the additional part $\delta_{\pm,r}^{(n)}$ is a Fourier integral operator of order zero with complex-valued symbol
\begin{gather*}
a_\pm^{(n)}(\xi):=\sum_{k=1}^n \frac{(i\xi)^{n+1}}{\big(i\xi\mp\frac{2\pi}{\beta}\big)^{n+1}}
\end{gather*}
and a complex-valued phase function
\begin{gather*}
\theta_\pm(x_\pm,\xi):=x_\pm\big(\xi\pm 2\pi i/\beta\big),
\end{gather*}
which is independent of $n$. In H\"ormander's terminology, i.e., where $x$ is chosen from a compact set $K\subset\R^n$, $\delta_{\pm,r}^{(n)}$ is a PsDO of order zero with the symbol,
\begin{gather*}
p(x_\pm,\xi):=\sum_{k=1}^n \frac{(i\xi)^{n+1}}{\big(i\xi\mp\frac{2\pi}{\beta}\big)^{n+1}}\;e^{-2\pi x_\pm}.
\end{gather*}
The generators $\delta_{\W_R,r}^{(n)}$, $\delta_{\V_+,r}^{(n)}$ and $\delta_{\D,r}^{(n)}$ with respect to the spacetime coordinates can be derived as in \eqref{delta0}. Thus, Assumption \ref{assumption-gen0} is proved by Theorem \ref{Borchers-Yngvason-gen} with
\begin{gather*}
\delta_0:=\delta_{\W_R}^{(0)},\;\delta_{\V_+}^{(0)},\;\delta_{\D}^{(0)}\quad\text{and}\quad\delta_r:=\delta_{\W_R,r}^{(n)},\;\delta_{\V_+,r}^{(n)},\;\delta_{\D,r}^{(n)}
\end{gather*}
for all $n\in\N$.

\vspace{8mm}

\hspace{-0,75cm}\textbf{ Acknowledgement:} The author is deeply grateful to K. Fredenhagen and M. Porrmann for helpful discussions.


\cleardoublepage

\addcontentsline{toc}{chapter}{\protect\numberline{Bibliography}}

\end{document}